\documentclass[preprint]{revtex4-2}

\usepackage{graphicx}
\usepackage{amssymb}
\usepackage{amsmath}
\usepackage{dcolumn}
\usepackage{bm}
\usepackage{epsfig}
\usepackage{color}
\usepackage{datetime}
\usepackage[T1]{fontenc}
\usepackage[utf8]{inputenc}

\newcommand{\bal}{\begin{align}}
\newcommand{\eal}{\end{align}}
\newcommand{\beq}{\begin{eqnarray}}
\newcommand{\eeq}{\end{eqnarray}}
\newcommand{\nneeq}{\nonumber \end{eqnarray}}

\newcommand{\nn}{\nonumber \\}
\newcommand{\es}{& = &}
\newcommand{\rs}{\, = \,}
\newcommand{\ps}{& + &}
\newcommand{\ms}{& - &}
\newcommand{\ts}{& \times &}

\newcommand{\nt}{\nn \ts}
\newcommand{\np}{\nn \ps}
\newcommand{\nm}{\nn \ms}

\newcommand{\cH}{ {\cal H} }

\newcommand{\cG}{ {\cal G} }

\newcommand{\cU}{ {\cal U} }

\newcommand{\2}{ \, { 1 \over 2} \, }

\begin{document}

\title{Elementary example of exact effective-Hamiltonian computation}
\author{ Stanis{\l}aw D. G{\l}azek }
\email{stglazek@fuw.edu.pl}
\affiliation{ Institute of Theoretical Physics\\ 
Faculty of Physics, University of Warsaw, Pasteura 5, 02-093
Warsaw, Poland }
\date{December 26, 2020 }

\begin{abstract}
We present an exact computation of effective Hamiltonians 
for an elementary model obtained from the Yukawa theory 
by going to the limit of bare fermions being infinitely heavy 
and bare bosons being at rest with respect to the fermions 
that emit or absorb them. The coupling constant can be 
arbitrarily large. The Hamiltonians are computed by solving 
the differential equation of the renormalization group procedure 
for effective particles (RGPEP). Physical fermions, defined in
the model as eigenstates of the effective Hamiltonians, are 
obtained in the form of an effective fermion dressed with a 
coherent state of effective bosons. The model computation 
illustrates the method that can be used in perturbative 
computations of effective Hamiltonians for realistic theories. 
It shows the mechanism by which the perturbative expansion 
and Tamm-Dancoff approximation increase in accuracy along 
the RGPEP evolution. 
\end{abstract} 

\maketitle


\section{Introduction}
\label{I}

Complexity of relativistic quantum field theory (QFT) 
implies a need for approximate computational methods.
One needs a systematic scheme for improving their accuracy. 
That is the case in computing observables using expansion 
in powers of a small coupling constant, solving eigenvalue 
problems using a limited basis in the space of states 
or using renormalization group methods. A combination 
of all three of these techniques requires a clear-cut 
pattern to follow. Such pattern can only be provided
by an exactly solvable model, because one needs the
exact solution to unambiguously assess accuracy of the 
approximate calculations. On the other hand, to obtain 
an exactly solvable model, one has to simplify a theory. 
A compromise needs to be struck between simplifying
and obtaining a helpful pattern. 

This article presents a novel, exact renormalization-group 
computation of effective Hamiltonians in a model that 
results from drastic but precisely specified simplifications 
of QFT, so that one can see the steps that would have to 
be reconstructed in an analogous computation of the effective 
Hamiltonians and their spectra in QFT. The presentation is 
thus quite limited but it includes enough of the QFT features 
for addressing the issues of high orders of perturbation theory, 
few-body approximations in the Fock space, 
renormalization-group improvements and the form of 
effective Hamiltonians that change, but not limit the 
number of interacting field quanta.

The computation presented in this article concerns 
a model that is obtained by drastically simplifying 
the Yukawa theory. The simplifications made here 
partly resemble the ones that Wilson adopted 
in formulating his approach to renormalization using a 
Yukawa-like model~\cite{Wilson1}, but they go much 
further. As a result, the ultraviolet divergences of a local 
theory are eliminated at the outset. This is useful because 
the goal of the presented computation is not to find the 
ultraviolet counter terms using the triangle of 
renormalization~\cite{triangle}, as it was in Wilson's 
case, but to deal with the issue of computation of an 
effective theory Hamiltonian after the right counter terms 
have already removed the divergences. The model 
computation includes a pattern of handling terms that 
are analogous to the finite parts of counter terms. 

It should be stressed that the history of models that 
incorporate elements of the Yukawa theory and are 
helpful in understanding renormalization in QFT 
beyond the weak-coupling expansion have a long 
history~\cite{HenleyThirring}. There are exactly 
solvable models among them, {\it e.g.}~\cite{Lee}. 
Also, a model may employ some elements of the 
Yukawa theory formalism and be exactly solvable 
without encountering any need for renormalization. 
For example, a class of two-level models for a system 
of a fixed number of fermions, whose Hamiltonians 
can be written using bilinear products of fermion creation 
and annihilation operators, could be solved exactly. 
One takes advantage of the SU(2) symmetry associated 
with the two levels~\cite{LMG,Debergh} or uses the 
symmetry of the model's boson representation~\cite{Providencia}. 
One can even show that such fermion systems exhibit 
thermalization when they are weakly coupled to a boson 
bath~\cite{Louw}. This variety of models that can be 
solved suggests to the author it should be clearly stated 
that the main purpose here is different. It is to apply 
a recently formulated renormalization group equation 
for Hamiltonians of QFT, to a simplified, one-level model 
for fermions coupled to bosons in a way that is analogous 
to the Yukawa theory coupling. In the model case, an exact 
operator solution to the equation is obtained in the form of 
a whole family of effective Hamiltonians that are strictly
equivalent. They all act in an infinite dimensional Fock 
space. Their common spectrum is obtained as a byproduct 
of the solution to the renormalization group equation.  
Comparisons and comments concerning the most similar 
models known to the author are provided in Sec.~\ref{effTD}.

The model used here is defined using the front form (FF) 
of Hamiltonian dynamics~\cite{Dirac} instead of the 
instant form (IF) used in~\cite{Wilson1}. It is known 
that to obtain the Wilson model from the front form 
of Yukawa theory one needs to consider the limit of 
fermions that are much heavier than the momentum 
cutoff parameter~\cite{GlazekPerry}, say $\Lambda$. 
Here, in addition, also the boson mass is assumed much 
greater than $\Lambda$. This limit is called the static 
limit, since bosons emitted or absorbed by fermions 
do not move with respect to their source. Further, the 
model we use does not include isospin, which leads to 
a significant simplification: only four distinct operators 
appear in the effective Hamiltonians. This feature will 
become clear in the course of computation. Despite 
these far reaching simplifications, the model interaction 
Hamiltonian changes the number of bosons and the 
number of Fock components involved in the dynamics 
is infinite. 

To compute the effective Hamiltonians, we use the method 
called the renormalization group procedure for effective 
particles (RGPEP). The RGPEP differs conceptually from the 
Wilson renormalization group procedure.  
Namely, instead of integrating out high-energy modes in the 
basis of the space of states in which the Hamiltonian acts, 
one changes the basis in the space of operators to which 
the Hamiltonian belongs. In other words, the Hamiltonian 
is seen as an element of the operator space formed by 
normal-ordered polynomials of bare creation and annihilation 
operators. The change of basis in the space of such polynomials 
is obtained by replacing the bare creation and annihilation 
operators with the analogous ones for the effective quanta of 
fields, called effective particles, see Sec.~\ref{solution} for details.
In perturbation theory, the effective particle operators are 
polynomials in terms of the bare particle operators and vice 
versa~\cite{algebra}. The interactions of effective particles 
are limited by the running cutoff $\Lambda$ that provides an 
upper bound on the magnitude of the invariant mass change 
that an interaction can cause. The RGPEP evolution of the 
computed Hamiltonians describes the variation of their form 
with the running cutoff.

The RGPEP employs the rules of the similarity renormalization 
group procedure for Hamiltonians~\cite{Similarity1,Similarity2} 
and takes advantage of the double-commutator feature of 
Wegner's flow equation for Hamiltonian matrices~\cite{Wegner}. 
In application to local QFT, the RGPEP has been recently 
illustrated in~\cite{Abelian}, which also includes references to 
the previous works. However, in all these examples one is forced 
to use the approximations that are not under precise control, such 
as the mentioned earlier weak-coupling expansion~\cite{triangle} 
or a limitation on a number of virtual particles, called the 
Tamm-Dancoff (TD) approximation~\cite{Tamm,Dancoff}. 
These approximations obscure the core features of the RGPEP in
the context of realistic theories. In contrast, the exact RGPEP 
computation of effective Hamiltonians in the model described 
here is quite transparent and the result has a clear interpretation 
in terms of the Fock-space image of physical states.

The RGPEP equation we solve, see Eq.~(\ref{dHdt}),  
determines the evolution of Hamiltonians using not the
cutoff parameter $\Lambda$ itself, but the parameter 
that is denoted by $t$ and corresponds to $\Lambda^{-2}$. 
Thus, $t$ varies from zero for the initial Hamiltonian to 
infinity for its final, diagonal form, in which all mass-changing 
interaction terms disappear. The ability to diagonalize Hamiltonians 
is the key design feature of the RGPEP equation. Quite 
generally, the design secures that the first-order solution 
of the RGPEP evolution equation results in vertex form 
factors whose width in momentum variables varies with 
$t$. The width tends to infinity or some cutoff value 
when $t \to 0$ and to zero when $t \to \infty$. In QFT, 
these form factors regulate singularities of the local 
interactions, {\it e.g.} see~\cite{Abelian}, and can be 
thought of as corresponding to a finite size of the effective 
particles. However, in the model solved here the situation 
is much simpler because of the static limit. One only 
obtains a running coupling constant, denoted by $g_t$, 
instead of a function of momentum, since the interacting 
particles are at rest with respect to each other. 

The paper is organized in the following way.
Section~\ref{model} describes the model Hamiltonian, 
{\it cf.}~\cite{HenleyThirring}. The model is derived in 
the FF of dynamics using the static limit of the Yukawa 
theory in Sec.~\ref{static}. It is rewritten in a more 
familiar energy notation of the IF of dynamics in 
Sec.~\ref{intuitive}. Section~\ref{solution} describes 
solution of the RGPEP equations. First the equations 
design is explained in Sec.~\ref{RGPEPdesign} and then
the discussion of solutions follows in Sec.~\ref{solution15}.
Operators that create and annihilate effective particles
are derived in Sec.~\ref{EP}, with final formulas in
Sec.~\ref{effectiveoperators}. Exact spectrum of the model
Hamiltonian in the Fock space representation is given
in Sec.~\ref{spectrum}. The issue of approximate 
computations is addressed in the remaining part of the 
article. Section~\ref{weakg} discusses the weak-coupling 
expansion. The TD approximation is discussed in 
Sec.\ref{TD}. Subsequently, Sec.~\ref{effTD}
introduces the concept of effective TD Hamiltonian 
matrices, including comments and comparisons to
related work on similar models. Section~\ref{comparison} 
briefly outlines the ways of comparing the model 
solution with realistic theories. Section~\ref{c} 
concludes the article and reviews motivation for
studies of QFT using the RGPEP.

\section{Model Hamiltonian}
\label{model}

The model Hamiltonian we consider is obtained from the
Yukawa theory using results of Ref.~\cite{GlazekPerry}. 
In that work, Eq.~(2.1) displays the canonical front-form 
Hamiltonian of Yukawa theory that has the structure
\beq
\label{HcYukawa}
H_c \es
H_f + H_b + H_{fb} + [{\rm other~terms}] \ ,
\eeq
where $H_f$ stands for the Hamiltonian of free fermions of mass $m$,
\beq
H_f \es \sum_\sigma \int[p] \, {m^2 + p^{\perp \, 2} \over p^+} \, b_{p\sigma}^\dagger b_{p\sigma} \ ,
\eeq
$H_b$ denotes a free Hamiltonian for bosons of mass $\mu$,
\beq
H_b \es \sum_\sigma \int[p] \, {\mu^2 + p^{\perp \, 2} \over p^+} \, a_p^\dagger a_p \ ,
\eeq
$H_{fb}$ is the fermion-boson interaction term,
\beq
H_{fb} 
\es 
g
\sum_{\sigma_1, \sigma_2} \int[p_1 p_2 p_3] \, \delta_{c.a} \, 
\bar u_1 \Gamma u_2 
\nt
b_{p_1\sigma_1}^\dagger (a_{p_3}^\dagger + a_{p_3})  b_{p_2\sigma_2} \ .
\eeq
The bracket $[{\rm other~terms}]$ indicates the terms 
that disappear in comparison with the first three in the limit 
of the fermion mass $m \to \infty$. The symbol $[p]$ 
conventionally denotes the measure $d^2p^\perp dp^+ 
\theta(p^+)/[2p^+(2\pi)^3]$ and $\delta_{a.c}$ is $2(2\pi)^3 
\delta^3(P_c - P_a)$, where $P_c$ and $P_a$ denote the 
total $+$ and $\perp$ momenta of the bare particles that 
are created and annihilated by the interaction, respectively. 
Fermion spinors are denoted by $u_1$ and $u_2$ and the 
matrix $\Gamma$ is set to 1. The coupling constant is 
denoted by $g$. Further model construction steps use only 
the first three terms in Eq.~(\ref{HcYukawa}).

\subsection{The static limit}
\label{static}

As a result of steps fully described in~\cite{GlazekPerry},
the Hamiltonian $H = H_f + H_b + H_{fb}$ is altered
in a way that leads to a formula resembling Eq.~(2.17) 
in that reference. One considers a fermion eigenstate of 
$H$ that carries an arbitrary momentum $P^+$ and 
$P^\perp$ and a fixed value of spin projection on $z$-axis. 
The state is a combination of the Fock component with 
one bare fermion in the same spin state and infinitely 
many Fock components each of which contains one bare 
fermion and some natural number of bare bosons. The 
fermion mass eigenvalue is written as $M = m +E$, 
where $E/m \ll 1$. Every boson kinematic momentum in 
a Fock component with $n$ bosons is parameterized 
according to the rule
\beq
p_{n,i}^+ \es y_{n,i} P^+ \ , \\
p_{n,i}^\perp \es y_{n,i} P^\perp + \kappa_{n,i}^\perp \ ,
\eeq
where $0 < i  \leq n$. The corresponding fermion 
momentum is given by 
\beq
p_n^+ \es x_n P^+ \ , \\
p_n^\perp \es x_n P^\perp - \kappa_{n,i}^\perp . . . - \kappa_{n,i}^\perp \ , 
\eeq 
where 
\beq
x_n \es 1 - y_{n,1}. . .- y_{n,n} \ .
\eeq
In the absence of bosons, the bare fermion carries 
the whole $P^+$ and $P^\perp$. Approximations
described in~\cite{GlazekPerry} are based on the
conditions that force $\kappa^\perp_{n,i} \ll m$ and 
$\sum_{i=1}^n y_{n,i} \ll 1$. A sufficient condition
is provided by imposing a cutoff that forces all
bosons to only have momenta relative to the fermion 
that are negligible in comparison with the fermion 
mass.

The interaction Hamiltonian $H_{fb}$ is supplied 
with a cutoff form factor, denoted below by $f_\Lambda$.
One assumes that $\Lambda \ll m$. The cutoff 
function enforces the condition $|P_c^2 - P_a^2| 
< \Lambda^2 \ll m^2$, where $P_c$ and $P_a$ 
denote the free total four-momenta of created and 
annihilated particles, respectively. In consequence,
all the fractions $y_i$ defined above tend to 0 
and the fermion fractions $x_n \to  1$. The resulting 
Hamiltonian that determines the mass eigenvalue 
$M$ for a physical fermion, see Eqs.~(2.16) and 
(2.17) in~\cite{GlazekPerry}, takes the form 
\beq
H_{\rm fermion} \es m b^\dagger b 
+ H_1 b^\dagger b \ ,
\eeq   
where $b$ denotes annihilation operator for a
bare fermion at rest and only one spin projection
on $z$-axis. The operator $H_1$ is the boson 
Hamiltonian associated with the states that contain 
one bare fermion,
\beq
\label{H1}
H_1 
\es
\int[q] \, \Big[ (1/2) \left( q^+ + {\mu^2 + q^{\perp \, 2} \over q^+} \right) 
\, a_q^\dagger a_q 
\np
g f_\Lambda(q) \,  (a_q^\dagger + a_q) \Big] \ .
\eeq
The boson momenta in all Fock sectors are identified 
according to the same relations $q^+=y_{n,i} m$ and $q^\perp 
= \kappa_{n,i}^\perp$. The half of the round bracket in Eq.~(\ref{H1})
equals energy of a boson with momentum $\vec q$ in which $2 q^z = q^+ - 
( \mu^2 + q^{\perp \, 2} )/ q^+$. It is possible to determine
the allowed momenta for bosons, including the sampling 
that Wilson adopted, by choosing the function 
$f_\Lambda(q)$. At this point one can further proceed
as in~\cite{GlazekPerry} and show that when the initial 
Yukawa theory includes isospin, the Hamiltonian one 
obtains in place of $H_1$ also includes isospin and 
matches the model Hamiltonian studied in~\cite{Wilson1}. 
In what follows a different path is taken. 

The next great simplification step we make here, which
was not made in~\cite{GlazekPerry}, is to assume 
that the boson mass $\mu$ is also much larger than 
the cutoff $\Lambda$. This assumption implies that 
the bosons cannot move with respect to the fermion 
that emits or absorbs them. In the limit $\Lambda/\mu
\to 0$, the Hamiltonian $H_1$ involves only bosons 
that are practically at rest relative to the fermion. 
One replaces all bosons nearly at rest with respect to 
the fermion by just one static boson mode, for which 
$q^\perp = 0$ and $q^+ = \mu$. We consider the 
model in which $\mu/m \to 0$. Finally, one can allow 
the static bosons to appear in the model also without 
a fermion. This way one arrives at the Hamiltonian of 
a model for which the exact effective Hamiltonians are 
computed using the RGPEP in the sections that follow.

\subsection{Intuitive notation}
\label{intuitive}

The model Hamiltonian introduced in the previous 
section is rewritten here using an intuitive notation that 
does not require familiarity with the front form of 
dynamics and instead relies on the intuition rooted
in the IF of dynamics,
\beq
\label{Hmodel}
H \es E_f \, b^\dagger b + E_b \, a^\dagger a 
+ g E_I \, b^\dagger (a^\dagger + a) b  \ , 
\eeq
where $E_f$, $E_b$ and $E_I$ are the fermion, 
boson and interaction energy parameters, $g$ 
is a coupling constant, while $b$ and $a$ are 
annihilation operators for the static fermion and 
boson, respectively. These operators and their 
hermitian conjugates are normalized to obey the 
standard (anti)commutation relations, of which
the only nonzero ones are
\beq
\{ b, b^\dagger\} \es 1  \ , \\
{[}a, a^\dagger{]} \es 1 \ .
\eeq
The Hamiltonian describes fermions of just
one spin state and preserves their number, 
which can only be 0 or 1. The number of 
bosons is neither specified nor limited and
it varies as a result of interactions.

In states with the fermion number equal zero, the 
interaction vanishes and the spectrum matches the 
one of a Hamiltonian for free bosons at rest, $H_b 
= E_b a^\dagger a$, with eigenvalues $E_{bn} = 
n E_b$, where $n$ is zero or a natural number. 
The corresponding normalized eigenstates of $H_b$ 
are $|n\rangle = (n!)^{-1/2} a^{\dagger n}|0\rangle$. 
In states with the fermion number equal 1, the 
Hamiltonian changes the boson number by 1 and 
the distribution of bosons needs to be computed. 

We apply the RGPEP to this model in the remaining 
part of this work. This means that instead of directly 
evaluating all of the Hamiltonian eigenvalues and 
eigenstates in terms of bare quanta, one introduces 
creation and annihilation operators for effective bosons 
and fermions and computes the effective Hamiltonians 
for them. The eigenstates of these effective Hamiltonians 
are then found in terms of the basis in the Fock space
that is constructed using the creation operators of the 
effective particles instead of the bare ones. The exercise 
is meant worth carrying out since one can unfold the 
simplifications used in deriving $H$ of Eq.~(\ref{Hmodel}) 
and look at the dynamics of Yukawa theory anew from 
the perspective of the model computation.

\section{Computation of the effective Hamiltonian}
\label{solution}

In the model considered here, the RGPEP equations 
for a family of renormalized Hamiltonians, labeled
by parameter $t$, can be written in the operator form,
\beq
\label{dHdt}
{d \over dt} \cH_t \es \left[\cG_t, \cH_t \right] \ , \\
\label{Gt}
\cG_t \es [H_f + H_b, \cH_t]  \ ,
\eeq
where $\cG_t$ is called the generator. The initial
condition at $t=0$ is provided by $H$ of Eq.~(\ref{Hmodel}),
which is denoted for that reason as $H_0$.

Equations~(\ref{dHdt}) and (\ref{Gt}) resemble 
Wegner's flow equations that describe the evolution 
of band-diagonal Hamiltonian matrices as functions 
of their width on energy scale; the width decreases 
as $t$ increases~\cite{Wegner}. There are two 
differences. One is that Eq.~(\ref{dHdt}) cannot be 
represented exactly by finite matrices, because the 
commutation relations for $a$ and $a^\dagger$ cannot.
The other one is that the generator $\cG_t$ is a commutator
of $\cH_t$ with the sum $H_f + H_b$ that does not 
depend on $t$, {\it cf.}~\cite{SzpigelPerry}. In 
the Wegner generator, the Hamiltonian matrix is 
commuted with its diagonal part that varies with $t$. 
It should be noted that Eq.~(\ref{dHdt}) is written 
for the opertaor $\cH_t$ that only contains 
$t$-independent creation and annihilation operators 
for bare particles, which are replaced by the corresponding 
$t$-dependent operators for effective particles in order 
to obtain the renormalized Hamiltonians $H_t$, see below. 

\subsection{Design of Eqs.~(\ref{dHdt}) and (\ref{Gt})}
\label{RGPEPdesign}

Design of Eqs.~(\ref{dHdt}) and (\ref{Gt}) originates 
in the idea that one can consider the Hamiltonian 
eigenvalue problems in local QFT in terms of some 
kind of effective quanta instead of the bare ones. 
The change from bare to effective quanta is motivated 
by the concept that the effective quanta interact in a 
so much less violent way than the bare quanta do that 
the eigenvalue problem may be convergent in the 
effective Fock-space basis, even if it does not exhibit 
convergence in the bare Fock-space basis. The appearance
of convergence is a consequence of the vertex factors 
that emerge in solutions of Eq.~(\ref{dHdt}). Emergence 
of such factors is the feature of double-commutator 
equations like Eq.~(\ref{dHdt}) with the generator
given by Eq.~(\ref{Gt}). The model application 
discussed here shows this feature in a simplified way,
see below, and demonstrates how convergence in the 
effective Fock-space basis improves with increase of $t$.

The key examples of physical elementary particle 
systems in terms of which one can think about the 
design of Eqs.~(\ref{dHdt}) and (\ref{Gt}) are hadrons. 
In QCD, represented in terms of bare quanta, hadrons 
are complex mixtures of infinitely many quarks and gluons 
that are confined. In the particle tables, most of the known 
hadrons are classified as bound states of just a few 
constituent quarks. The design of the RGPEP equations 
can be described as aiming at the derivation of a 
mathematically precise connection between these two 
pictures of hadrons. 

The creation and annihilation operators for effective 
particles are defined using a unitary transformation
of the form
\beq
\label{at}
q_t \es \cU_t \, q \, \cU^\dagger_t \ ,
\eeq
where $q$ stands for the operators $a$, $a^\dagger$, 
$b$ or $b^\dagger$, and 
\beq
\label{cUt}
\cU_t^\dagger \es T \exp \left(\int_0^t d\tau \, \cG_\tau \right) \ .
\eeq
The symbol $T$ denotes ordering in $\tau$. 
The Hamiltonian operator
\beq
\label{Ht}
H_t \es \cU_t \, \cH_t \, \cU^\dagger_t
\eeq
is defined to be the same as the initial one, $H_t = H_0$, 
but $H_t$ is expressed in terms of the effective particle 
operators $a_t$, $a_t^\dagger$, $b_t$ and $b_t^\dagger$ 
instead of the initial operators $a$, $a^\dagger$, $b$ and 
$b^\dagger$ that correspond to $t=0$. Thus, in $H_t$, 
the coefficients of products of the effective creation and 
annihilation operators are different from the coefficients of 
products of the corresponding initial operators in $H_0$. 
The coefficients in $H_t$ contain factors that follow from 
the double-commutator structure of Eq.~(\ref{dHdt}). 
These factors are obtained in the process of solving 
Eq.~(\ref{dHdt}). They emerge in a way similar to the 
emergence of the band-diagonal matrices from the 
Wegner flow equation.

If the model were divergent, as it is the case for bare 
Hamiltonians in local QFT, $H_0$ would be supplied 
with the counter terms that would be computed from 
the condition that the coefficients of effective particle 
operators in $H_t$ for any finite, fixed value of $t$ 
are not sensitive to the adopted regularization of the 
divergences. Since the model Hamiltonian of 
Eq.~(\ref{Hmodel}) does not generate divergences, 
the computation of counter terms to divergent expressions 
is not needed and this aspect of local QFT is not illustrated 
in the model solution. The divergence counter term computation 
in QFT significantly complicates the RGPEP procedure with 
a lot of details that depend on the adopted regularization. 
These largely arbitrary details obstruct the conceptual view 
of the method while the model computation makes it clear.
Counter terms appear in the model computation only in a 
finite form, which is analogous to the appearance of the 
unknown finite parts of the divergence counter terms in 
local QFT.

\subsection{Solution of Eq.~(\ref{dHdt})}
\label{solution15}

In order to solve Eq.~(\ref{dHdt}), one writes 
\beq
\cH_t \es (E_f + \delta E_{ft}) b^\dagger b 
+ E_b a^\dagger a 
\np
 g_t E_I \, b^\dagger (a^\dagger + a) b \ , 
\eeq
where the subscript $t$ indicates dependence
on that argument. Only four distinct Fock-space 
operators appear in this formula because no 
other operators are generated from the initial 
condition of Eq.~(\ref{Hmodel}). Using a dot to 
indicate the derivative, one obtains Eq.~(\ref{dHdt}) 
in the form
\beq
\label{dot1}
\delta \dot E_{ft} b^\dagger b &+& 
\dot g_t E_I \, b^\dagger (a^\dagger + a) b \nn
\es
[ \cG_t, 
(E_f + \delta E_{ft}) b^\dagger b + E_b a^\dagger a 
\np
g_t E_I \, b^\dagger (a^\dagger + a) b ] \ , \\
\label{cGt}
\cG_t \es g_t E_b E_I \, b^\dagger (a^\dagger - a) b \ .
\eeq
The generator takes the simple form since 
the fermion number is conserved by the 
interaction. Evaluation of the commutator on the right-hand side
of Eq.~(\ref{dot1}) yields
\beq
\label{dot2}
\delta \dot E_{ft} \, b^\dagger b &+& 
\dot g_t E_I \, b^\dagger (a^\dagger + a) b \nn
\es
-g_t E_b^2 E_I \, b^\dagger (a^\dagger + a) b
\nm
2 g_t^2 E_b E_I^2 \, b^\dagger b \ .
\eeq
Equating coefficients in front of the same operators
on both sides, one gets
\beq
\label{dot3}
\delta \dot E_{ft} 
\es 
-2 g_t^2 E_b E_I^2 \ , \\
\label{dot4}
\dot g_t \es
-g_t E_b^2 \ .
\eeq
These are ordinary differential equations and solving
them leads to the solution of Eq.~(\ref{dHdt}) 
in the form
\beq
\label{solutioncH}
\cH_t \es 
\left[ E_f + g_t^2 \Delta_t \right] \, b^\dagger b 
+ 
E_b \, a^\dagger a 
\np
g_t E_I \, b^\dagger (a^\dagger + a) b \ ,
\eeq
where 
\beq
\label{gt}
g_t \es g e^{- E_b^2 t} \ , \\
\label{Deltat}
\Delta_t \es (1 - e^{2E_b^2t})E_I^2/E_b \ .
\eeq
This result shows that the increase of $t$ from zero to infinity
causes the effective fermion-boson coupling constant $g_t$ 
to decrease exponentially fast from its initial value $g$ to zero
at the rate given by an inverse of the boson energy squared. 
This is the promised suppression of interactions by the vertex 
factor. One obtains the vertex factor in this model solely in 
the form of a varying coupling constant $g_t$ instead a whole 
form factor that is a function of momentum and energy transfer 
between quanta in the vertex. The simplification occurs because 
the model contains only static modes for fermions and bosons. 

The boson energy $E_b$ stays constant as a function of $t$. 
The fermion energy, $E_f + g_t^2 \Delta_t$, evolves from 
the initial value $E_f$ to the final fermion eigenvalue energy 
\beq
E_{f\infty} \es \lim_{t \to \infty}
E_f + g_t^2 \Delta_t \rs 
E_f - g^2 E_I^2/E_b \ .
\eeq
It seems that $E_{f\infty}$
may be negative. However, it could only happen outside the 
range of approximations made in the model, where the 
fermion energy $E_f$ is assumed much larger than the boson 
energy $E_b$ and much larger than the energy change due 
to the interaction, $g E_I$, while $E_b$ and $E_I$ are of
similar magnitude. Therefore, for any fixed value of $g$, 
one only considers $E_f$ much larger that $g^2 E_I^2/E_b$.

\section{Effective particles}
\label{EP}

Solution for the operator $\cH_t$ in Eq.~(\ref{solutioncH})
is transformed into the Hamiltonian for effective particles
using the operator $\cU_t$ according to Eq.~(\ref{Ht}).
The result is
\beq
\label{Hteff}
H_t \es 
\left( E_f + g_t^2 \Delta_t \right)\, b_t^\dagger b_t 
+ 
E_b \, a_t^\dagger a_t 
\np
g_t E_I \, b_t^\dagger (a_t^\dagger + a_t) b_t \ ,
\eeq
where $b_t$ and $a_t$ are given by Eq.~(\ref{at}). 
Knowing $\cG_t$ in Eq.~(\ref{cGt}), one obtains
from Eq.~(\ref{cUt}) that
\beq
\label{cUt1}
\cU_t^\dagger 
\es 
e^{c_t \,  b^\dagger (a^\dagger - a) b} 
\rs
1 + \left[ e^{c_t \,  (a^\dagger - a)} - 1 \right] \ b^\dagger  b \ ,
\eeq
where 
\beq
\label{ct}
c_t \es  ( g - g_t ) E_I/E_b \ .
\eeq
Therefore,
\beq
\label{at1}
a_t \es a(1-b_t^\dagger b_t) + b_t^\dagger \, a \,  b_t \ , \\
\label{bt1}
b_t \es   e^{ c_t \, (a^\dagger - a) } b \ .
\eeq
Analogous formulas hold for creation operators
$a_t^\dagger$ and $b_t^\dagger$, obtained
by hermitian conjugation.

\subsection{Effective particle operators}
\label{effectiveoperators}

It is visible in Eq.~(\ref{at1}) that the effective boson
operators $a_t$ are equivalent to the bare ones in 
the subspace of Fock space without effective fermions, 
for in that case $b_t \equiv 0$. In the 
subspace that contains one effective fermion, one 
has $b_t^\dagger b_t \equiv 1$ and is left with
\beq
\label{at3}
a_{t1} \es b_t^\dagger \, a \,  b_t \ ,
\eeq
and a corresponding relation for $a_{t1}^\dagger$.
Evaluation yields
\beq
a_{t1} \es (a + c_t) \, b^\dagger b \ ,
\eeq
and $a_{t1}^\dagger$ is obtained by conjugation.

In summary, the annihilation operator for effective 
fermion, $b_t$, is given by Eq.~(\ref{bt1}), and  
the annihilation operator for an effective boson is 
\beq
\label{atfinal}
a_t \es a +  c_t \, b^\dagger b \ ,
\eeq
where $c_t$ is given by Eq.~(\ref{ct}). The corresponding
creation operators are obtained by hermitian conjugation. 
Using these results, one can check by a direct calculation 
that the effective Hamiltonian $H_t$ of Eq.~(\ref{Hteff}) 
is equal to the initial Hamiltonian $H = H_0$ of Eq.~(\ref{Hmodel}).

\section{Exact spectrum in the Fock space}
\label{spectrum}

One observes that there are three ways of seeking 
the model Hamiltonian spectrum. In the first way, 
one uses the Hamiltonian expressed in terms of the 
initial particle operators that correspond to $t=0$.
In the second way, one uses the Hamiltonian 
expressed in terms of effective particle operators
for some finite value of the RGPEP parameter $t$. 
The third way is reduced to inspection of the
effective Hamiltonian with $t=\infty$. The respective
forms of one and the same Hamiltonian $H=H_0$ 
of Eq.~(\ref{Hmodel}) are
\beq
\label{Hmodel1}
H_0 \es E_f \, b^\dagger b + E_b \, a^\dagger a 
+ gE_I \, b^\dagger (a^\dagger + a) b  \ , \\
\label{Hteff1}
H_t \es 
\left( E_f + g_t^2  \Delta_t  \right) \, b_t^\dagger b_t 
+ 
E_b \, a_t^\dagger a_t 
\np
g_t E_I \, b_t^\dagger (a_t^\dagger + a_t) b_t \ , \\
\label{Hinfty1}
H_\infty \es 
E_{\rm fermion} \, b_\infty^\dagger b_\infty 
+ 
E_b \, a_\infty^\dagger a_\infty \ .
\eeq
where $\Delta_t$ is given in Eq.~(\ref{Deltat}).
Taking into account the commutation relations that the 
operators with $t=\infty$ obey, one sees that the 
eigenvalues are
\beq
\label{fermioneigenvalue}
E_{\rm fermion} \es 
\label{nbosonseigenvalue}
\lim_{t \to \infty} \left( E_f + g_t^2 \Delta_t \right) \\ 
\es E_f - g^2E_I^2/E_b \ , \\
E_{n \, \rm bosons} \es n E_b \ , \\
\label{fermionbosoneigenvalues}
E_{{\rm fermion} + n \, \rm bosons} \es E_{\rm fermion} + n E_b \ ,
\eeq
and the corresponding normalized eigenstates are
\beq
\label{fermionfinal}
| {\rm fermion} \rangle \es b_\infty^\dagger |0\rangle \ , \\
\label{bosonsfinal}
|n \, {\rm bosons} \rangle \es {1 \over \sqrt{n!}} a_\infty^{\dagger n} |0\rangle \ , \\
\label{fermionbosonsfinal}
|{\rm fermion}+n \, {\rm bosons} \rangle \es {1 \over \sqrt{n!}} a_\infty^{\dagger n} b_\infty^\dagger |0\rangle \ ,
\eeq
where $|0\rangle$ denotes the model Hamiltonian 
ground state that contains no physical particles.
According to Eqs.~(\ref{at1}) and (\ref{bt1}),
\beq
\label{binfty}
b_\infty \es   e^{ g (E_I/E_b) \, (a^\dagger - a) } b \ , \\
\label{ainfty}
a_\infty \es a +  g (E_I/E_b) \, b^\dagger b \ .
\eeq
A physical fermion state is composed of the bare 
fermion and a coherent state of bosons. 
Since $a^\dagger - a = a_t^\dagger - a_t = 
a_\infty^\dagger - a_\infty $,
one can speak of the coherent state of bare as
well as effective or physical bosons. The $n$-boson 
eigenstates without a fermion are the same as if 
the interaction were absent. 

\section{Weak-coupling expansion}
\label{weakg}

In the weak-coupling expansion one hopes to 
gain some insight concerning solutions of a 
theory assuming that the coupling constant 
in the interaction terms is  a very small number. 
After evaluating some quantity of interest using 
expansion in powers of an infinitesimal coupling, 
one can check how large the coupling would 
have to be for the result to match data. Then 
there comes the question of how large the 
remaining terms in the expansion are.

In the model with the coupling constant $g$ not 
very small, such procedure is not viable as an 
approximation method for obtaining eigenstates 
of the Hamiltonian $H$ in terms of bare particle 
operators that appear in its form $H_0$. This form 
corresponds to the Yukawa theory expressed in 
terms of bare degrees of freedom. Although the 
fermion eigenvalue $E_{f\infty}$ is just a quadratic 
function of $g$ and one might hope that an expansion 
up to terms order $g^2$ may be sufficient, the 
fermion eigenstate contains terms with all powers 
of the product $g$ times the bare boson creation 
operator acting on the vacuum state. 

Quite different situation is encountered when 
one uses the Hamiltonian in its form $H_t$ 
with $E_b^2 t$ sufficiently large for $g_t$ of 
Eq.~(\ref{gt}) to be small. The eigenstates 
without a fermion are just free effective bosons 
created by $a_t^\dagger$ from the vacuum state. 
The eigenstates with a fermion are given by 
$b_t^\dagger |0\rangle$ plus admixtures of 
effective bosons that are created from the fermion 
state with strength $g_t$ instead of $g$. One 
sees in Eq.~(\ref{gt}) that $g_t$ can be small 
for arbitrarily large $g$ when $t$ is made sufficiently 
large. In that case, the fermion state can be 
approximated well by using the expansion in 
powers of $g_t$. 

The mechanism described above can be illustrated by the 
perturbative expansion up to second-order for the fermion 
energy eigenvalue and the corresponding eigenstate. In 
general, a perturbative expansion is obtained by writing
\beq
|\psi \rangle \es
(\psi_{00} + \psi_{01} + \psi_{02} +...) \, b_t^\dagger|0\rangle
\np
(\psi_{10} + \psi_{11} + \psi_{12} +...) \, a_t^\dagger b_t^\dagger|0\rangle
\np
(\psi_{20} + \psi_{21} + \psi_{22} +...) \, a_t^\dagger a_t^\dagger b_t^\dagger|0\rangle 
+ . . . \ ,
\eeq
where $\psi_{mn} \sim g_t^n$. The eigenvalue problem reads
\beq
\label{eigenproblemPT}
H_t |\psi \rangle \es  (E_0 + E_1 + E_2 + . . . ) |\psi \rangle \ ,
\eeq
where $E_n$ is of order $g_t^n$. Assuming that the 
dominant coefficient in front of $b_t^\dagger |0\rangle$ 
is $\psi_{00}$ of order 1, one can limit the effective Fock-space 
expansion to only three terms:  one effective fermiom, one 
effective fermion and one effective boson, and one effective 
fermion and two effective bosons. Coefficients of the components 
with more effective particles are of order $g_t^n$ with $n > 2$. 
By projecting both sides of Eq.~(\ref{eigenproblemPT}) on these 
three basis states, one obtains three equations. Projection on 
the component $b_t^\dagger |0\rangle$ yields
\beq
0 
\es
\left( E_f + g_t^2 \Delta_t - E_0 - E_1 - E_2 -. . .\right)
\nt
(\psi_{00} + \psi_{01} + \psi_{02} +...) 
\np
g_t E_I (\psi_{10} + \psi_{11} + \psi_{12} +...) \ .
\eeq
Projection on $a_t^\dagger b_t^\dagger |0\rangle$ leads to 
\beq
0
\es 
\left( E_f + g_t^2 \Delta_t - E_0 - E_1 - E_2 -. . . \right)
\nt
(\psi_{10} + \psi_{11} + \psi_{12} +...) 
\np
E_b(\psi_{10} + \psi_{11} + \psi_{12} + . . .) 
\np
g_t E_I (\psi_{00} + 2 \psi_{20} + \psi_{01} + 2 \psi_{21} + . . .) \ .
\eeq
Projection on $a_t^\dagger a_t^\dagger b_t^\dagger |0\rangle$ produces
\beq
0
\es 
2 \left( E_f + g_t^2 \Delta_t - E_0 - E_1 - E_2 -. . . \right)
\nt
(\psi_{20} + \psi_{21} + \psi_{22} +. . .) 
\np
4 E_b(\psi_{20} + \psi_{21} + \psi_{22} +. . .) 
\np
2 g_t E_I [ \psi_{10} + \psi_{11} + \psi_{12}+ . . .] \ .
\eeq
Each of these equations contains terms proportional 
to powers of $g_t$. Equating coefficients of 1, $g_t$ 
and $g_t^2$ on both sides of these equations, one 
arrives at a set of 9 equations that must be satisfied 
simultaneously. Assuming that $\psi_{00} = 1$, 
setting $\psi_{01} = \psi_{02} = 0$ and introducing 
the normalization factor $N$,  one obtains 
\beq
E \es E_f - {1 \over E_b} \left( g_t e^{E_b^2 t} E_I \right)^2 \ , \\
|\psi \rangle \es
N
\left[   b_t^\dagger|0\rangle  
      - {1 \over E_b} g_t E_I \, a_t^\dagger b_t^\dagger|0\rangle 
\right.
      \np
\left.
 \2 \left( {1 \over E_b} g_t E_I \right)^2  \, a_t^\dagger a_t^\dagger b_t^\dagger|0\rangle \right]  \ .
\eeq
The term $(g_t E_I)^2/E_b$ in the effective fermion energy 
in Eq.~(\ref{Hteff1}) cancels the second-order self-interaction 
term that results from emission and absorption of an effective 
boson. Thus, even though in the model the fermion self-interaction 
is finite, the term $(g_t E_I)^2/E_b$ in the effective fermion 
energy term in the Hamiltonian $H_t$ appears in the role of 
a finite part of the fermion self-interaction counter term
when the parameter $t$ tends to zero and its inverse plays 
the role of a cutoff $\Lambda^2$. The finite part is positive, 
vanishes when $\Lambda \to 0$ or $t \to \infty$ and implies 
that in that limit the effective fermion energy in $H_t$ approaches 
the physical fermion eigenvalue $E_{\rm fermion}$. One could 
replace the effective fermion energy term in $H_t$ by the 
eigenvalue and ignore the self-interaction effects.

\section{The Tamm-Dancoff approximation}
\label{TD}

The idea of the TD approximation~\cite{Tamm,Dancoff} 
is to limit the Hamiltonian eigenvalue problem to a subspace of 
the Fock space defined by a limit on the number of virtual particles.
One assumes that the eigenstate components with more particles 
than the limiting number have a small probability and can be neglected 
in the first approximation. Such approach was also proposed in the 
context of solving QCD in the front form of Hamiltonian dynamics, 
using the idea that a suitable renormalization group algorithm, including 
the Fock-sector dependent counter terms, could be used to identify the 
dominant features of the dynamics as the limit on the number of particles 
is increased. Subsequently, one could attempt to compute corrections to 
the dominant picture using the methods of perturbative expansion and 
successive approximations~\cite{LFTammDancoff}, including some 
form of the coupling coherence~\cite{PerryWilson}.

\begin{figure}[ht!]
          \caption{Expectation value of the number of effective-bosons, 
                      see Eq.~(\ref{<Nt>}), in the physical fermion eigenstate 
                      of Eq.~(\ref{fermionfinal}) as a function of the RGPEP 
                      scale parameter $t$. The three curves correspond to the 
                      three values 2, 1 and 1/2 of the coupling constant $g_0$ 
                      in Eq.~(\ref{<Nt>3}), defined as the effective coupling 
                      constant $g_t$ for $t$ equal $t_0 = 1/E_b^2$, assuming 
                      that the free boson energy $E_b$ equals the fermion-boson 
                      interaction energy parameter $E_I$ in the model Hamiltonian, 
                      see Sec.~\ref{TD}. It is visible that the TD approximation  
                      becomes increasingly accurate when $t$ increases, since 
                      $\langle N_t \rangle$ decreases exponentially fast with 
                      increase of $t$.} 
          \label{fig:TDillustration}
\begin{center}
         \includegraphics[width=.4\textwidth]{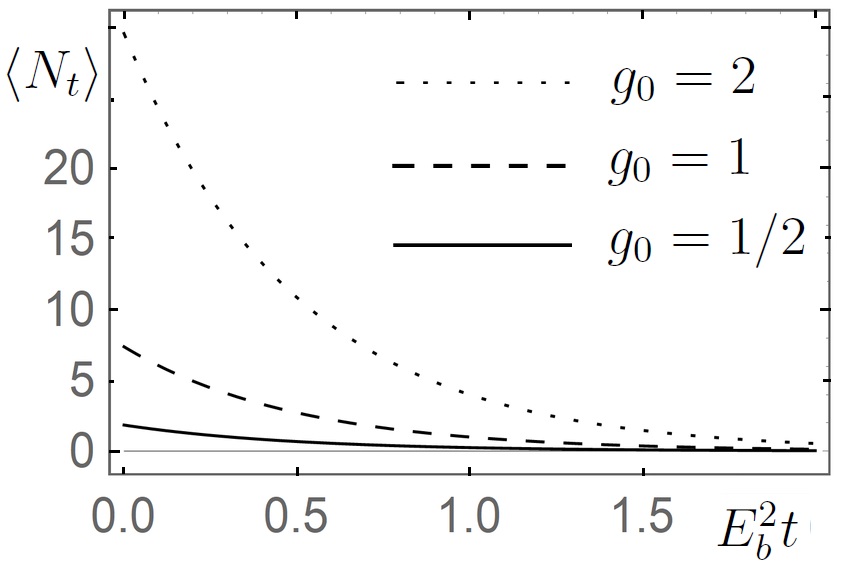}
\end{center}
          \end{figure}

In case of the RGPEP, the key feature that influences the accuracy 
of the TD type of approach to realistic theories is that instead of 
the bare, original field quanta one limits the number of the effective 
quanta. The idea is illustrated using Fig.~\ref{fig:TDillustration}. 
It shows plots of the expected number of virtual effective bosons in 
the physical fermion state as a function of the RGPEP scale parameter 
$t$. The plotted value is defined by 
\beq
\label{<Nt>}
\langle N_t \rangle
\es
\langle {\rm fermion}| a_t^\dagger a_t |{\rm fermion} \rangle \ ,
\eeq
where the fermion state is given in Eq.~(\ref{fermionfinal}).
Using Eqs.~(\ref{atfinal}) and (\ref{binfty}) one obtains
\beq
\label{<Nt>2}
\langle N_t \rangle \es g_t^2 (E_I/E_b)^2 \ ,
\eeq
which for $E_I=E_b$ yields the expected number of virtual
effective bosons in a physical fermion,  
\beq
\label{<Nt>3}
\langle N_t \rangle \es g_t^2 
\rs
g_0^2 \, e^{2 E_b^2 (t_0-t)} \ .
\eeq
The coupling constant $g_0$ is the value that $g_t$
takes when $t=t_0$. We set the value of $t_0$ to 
$E_b^{-2}$, since this value of the running cutoff 
corresponds in magnitude to the energy change 
associated with emission or absorption of just one 
boson. The value of $g_0$ is arbitrary. To provide 
examples of the numbers involved, three values of 
the coupling constant $g_0$ are arbitrarily selected: 
2, 1 and 1/2. The three curves shown in 
Fig.~\ref{fig:TDillustration} correspond to these 
three values of $g_0$. The number of virtual bosons 
in a physical fermion strongly depends on the value 
of $g_0$ and these three values are sufficient to 
illustrate the dependence. Each of the chosen values 
corresponds to a different value of the bare coupling 
constant $g$ in Eq.~(\ref{Hmodel}), $g = e g_0$. 

It is visible in Fig.~\ref{fig:TDillustration} that approximations 
of the TD type with just one or two virtual bosons 
do not apply in terms of the bare particles if the coupling constant
$g_0$ is not sufficiently small. For example, if $g_0=2$, the 
expectation value $\langle N_0 \rangle$ is almost 30. However, 
when $t$ grows, the expectation value $\langle N_t \rangle$ 
decreases. In the model, where one possesses the exact solution 
to the RGPEP equation, Fig.~\ref{fig:TDillustration} shows that 
for $t$ exceeding $E_b^{-2}$ the effective interaction vertex 
suppression factor can become so small that the strength of the 
bare coupling constant is overcome and the TD approximation 
represents the physical fermion accurately in terms of a small 
number of the corresponding virtual effective particles.

\section{TD Hamiltonian matrices}
\label{effTD}

If the coupling constant $g_t$ is sufficiently 
small and the parameter $t$ large enough for 
the RGPEP form factors to suppress the interaction 
terms in $H_t$ that change the number of effective 
particles, then the TD approximation may be valid.
In that case one can define the effective Hamiltonian
matrices that describe the dynamics in terms of a 
limited number of effective Fock-space basis states. 
We call them the effective TD Hamiltonian matrices, 
or just TD matrices, denoted by $H_{\rm TD \, t}$. 
One can compute them following the pattern illustrated 
below in terms of our model. 

Consider the Hamiltonian $H_t$ in which the effective, 
particle number-changing interaction term is weak 
enough to expect that the TD approximation is 
reasonable. Suppose one is interested in an approximate 
computation of observables for a physical fermion.
In the model, the physical fermion is known exactly.
It is represented by the state $|{\rm fermion}\rangle$ 
in Eq.~(\ref{fermionfinal}). However, in an approximate
calculation in a realistic theory a physical fermion 
state would not be known exactly.

Suppose one expects that the physical fermion state 
is dominated by the basis state $b_t^\dagger|0\rangle$,
while the basis state $a_t^\dagger b_t^\dagger|0\rangle$ 
provides the leading correction. Still smaller corrections 
involve the basis states $a_t^{\dagger \, n} b_t^\dagger
|0\rangle $ with $n>1$. To describe the physical fermion 
state using the TD approximation, one computes the 
matrix $H_{\rm TD \, t}$ that acts on the coordinates
of states in the subspace of the Fock space that is spanned 
by the basis states with one effective fermion and a limited 
number of effective bosons. If instead of the physical 
fermion one were interested in the properties of states 
$|{\rm fermion}+n_b \, {\rm bosons} \rangle$, one would 
first compute Hamiltonian matrix $H_{\rm TD \, t}$ that 
acts on the coordinates in the subspace spanned by the 
effective basis states $a_t^{\dagger \, n_b-1}b_t^\dagger
|0\rangle$, $a_t^{\dagger \, n_b   }b_t^\dagger|0\rangle$ 
and $a_t^{\dagger \, n_b+1}b_t^\dagger|0\rangle$.
Corrections would follow from enlarging the matrix to
include coordinates in directions of basis states with
$n_b \pm 2$ effective bosons, {\it etc.} 

A simple illustration of the TD approximation is 
obtained in case of the physical fermion and 
the assumption that the matrix $H_{\rm TD \, t}$ 
only acts on the two-dimensional vectors of 
coordinates in the Fock-subspace spanned by 
the basis states $b_t^\dagger |0\rangle$ and
$a_t^\dagger b_t^\dagger|0\rangle$. The first 
approximation is obtained by writing 
\beq
\label{fTD}
|{\rm fermion_{TD}}\rangle 
\es
x_{t0} \, b_t^\dagger|0\rangle
+
x_{t1} \, a_t^\dagger b_t^\dagger|0\rangle
\np
{1 \over \sqrt{2}} x_{t2} \, a_t^{\dagger \,2} b_t^\dagger|0\rangle \ .
\eeq
Then one observes that the physical fermion eigenvalue 
probem has the form  
\beq
H_t \, |\psi \rangle   \es 
E_{\rm TD} \, |\psi \rangle  \ , \\
|\psi \rangle 
\es 
|{\rm fermion_{TD}}\rangle + |n_b > 2\rangle \ ,
\eeq
where $|n_b > 2\rangle$ stands for all components with 
more effective bosons than 2. Projecting this equation 
on the same range of components that appears in
Eq.~(\ref{fTD}), one obtains the matrix equation 
\beq
\label{TD1}
\left[
\begin{array}{ccc}
h_{2,2} & h_{2,1} & 0         \\
h_{1,2} & h_{1,1} & h_{1,0} \\
0         & h_{0,1} & h_{0,0}
\end{array}
\right]
\left[
\begin{array}{c}
x_{t2}  \\
x_{t1}  \\
x_{t0}
\end{array}
\right]
&+&
\left[
\begin{array}{c}
h_{2,n_b>2} \\
0  \\
0 
\end{array}
\right] \nn
\es
E_{\rm TD} 
\left[
\begin{array}{c}
x_{t2}  \\
x_{t1}  \\
x_{t0}
\end{array}
\right] \ ,
\eeq
where the matrix elements are
\beq
h_{m,n} \es {1 \over \sqrt{ m! n!} } \langle 0| b_t a_t^m \, H_t a_t^{\dagger n} b_t^\dagger |0\rangle
\eeq 
with $0 \leq m,n \leq 2$. The TD approximation 
amounts to setting $h_{2,n_b>2} = 0$. The 
effective TD Hamiltonian matrix is defined by
\beq
H_{\rm TD \, t \, mn } \es h_{m,n} \ .
\eeq
Its eigenvalue problem reads
\beq
\sum_{n=0}^2 H_{\rm TD \, t \, mn } \ x_n
\es
E_{TD} \ x_m \ .
\eeq
One has 
\beq
H_{\rm TD \, t \, 22 } \es E_f + g_t^2  \Delta_t  + 2 E_b   \ , \\
H_{\rm TD \, t \, 21 } \es H_{\rm TD \, 12 \, t} 
                              \rs  \sqrt{2} g_t E_I  \ , \\
H_{\rm TD \, t \, 11 } \es E_f + g_t^2  \Delta_t  + E_b  \ , \\
H_{\rm TD \, t \, 10 } \es H_{\rm TD \, 01 \, t} 
                              \rs g_t E_I \ , \\
H_{\rm TD \, t \, 00 } \es E_f + g_t^2  \Delta_t \ .
\eeq
The eigenvalues $E$ written in the form $E = E_f + g_t^2  \Delta_t + x E_b$
obey the equation 
\beq
(2-x)(1-x) x + \alpha (2 - 3x) \es 0 \ ,
\eeq
where $\alpha = (g_t E_I/E_b)^2$. If $\alpha$ 
were zero due to $g_t=0$, the three eigenvalues 
$E_n = E_f + g_t^2  \Delta_t + x_nE_b$ with
$x_n = n$ would correspond to a free effective 
fermion and $n$ free bosons. Assuming that $x 
= n + y \alpha$ and neglecting higher powers of 
$\alpha$ one obtains $E_n = E_{\rm fermion} + 
n E_b$ for $n$ equal 0 or 1, as expected 
on the basis of the exact solution given in 
Eq.~(\ref{fermionbosoneigenvalues})
and Eqs.~(\ref{fermionfinal}) or (\ref{fermionbosonsfinal}), 
respectively. 
Higher order terms in the expansion of $x$ in 
powers of $\alpha$ can be used to compare the 
TD approximation with the weak-coupling expansion. 
Next level of the TD approximation would be obtained 
by introducing the component $a_t^{\dagger \, 3} 
b_t^\dagger |0\rangle /\sqrt{6}$ and neglecting 
$h_{2,n_b>3}$. 

The coordinates $x_n$ in realistic theories would 
not be just numbers but unknown functions of only
$3n$ relative momentum variables and discrete 
quantum numbers of fermions and bosons. The 
number of momentum arguments would be the 
same in the non-relativistic and relativistic theories 
because the total momentum of the eigenstates 
drop out from the TD matrix problem and the 
eigenvalues $E$ are solely the masses squared of 
the physical systems.

One can use Eq.~(\ref{TD1}), ignoring $h_{2,n_b>2}$,
to evaluate the Fock-space coordinate $x_{t2}$ in terms 
of the coordinates $x_{t1}$ and $x_{t0}$ using a fully 
non-perturbtive Gaussian elimination or the so-called 
$R$ operation, the latter when either the boson energy 
$E_b$ is large~\cite{Wilson2} or $g_t$ is small. The 
Gaussian elimination yields
\beq
x_{t2} \es {1 \over E_{TD} - h_{2,2}} h_{2,1}  \, x_{t1} \ ,
\eeq
which can be put into the remaining two equations.
The result is 
\beq
\label{TD2}
&&
\left[
\begin{array}{cc}
  h_{1,1} + h_{1,2} {1 \over E_{TD} - h_{2,2}} h_{2,1} & h_{1,0} \\
  h_{0,1} & h_{0,0}
\end{array}
\right]
\left[
\begin{array}{c}
x_{t1}  \\
x_{t0}
\end{array}
\right] \\
\es
E_{\rm TD} 
\left[
\begin{array}{c}
x_{t1}  \\
x_{t0}
\end{array}
\right] \ .
\eeq
This is the TD matrix eigenvalue problem including 
the fermion self-interaction in the fermion-boson 
component. Note that the eigenvalue $E_{TD}$
appears on both sides of the problem, which requires
a non-perturbative matching of its left-hand side
value with its value on the right-hand side. If instead 
of the Gaussian elimination one used the operation 
$R$~\cite{Wilson2} in expansion up to second power 
of the coupling constant $g_t$, including the perturbative 
orthogonality and normalization corrections, then the 
eigenvalue $E_{TD}$ in Eq.~(\ref{TD2}) on the 
left-hand side would be replaced by $E_f + E_b$.
Even though in this case the left-hand side matrix 
would only contain terms of order up to $g_t^2$, 
the eigenstates would depend on $g_t$ in a way 
specific to the particular TD approximation. The 
issue would then be what changes occur when one 
attempts to improve the approximation by including 
more effective particles or higher powers of $g_t$ 
in the TD Hamiltonian matrices. An example of a 
phenomenological study based on the hypothesis 
that gauge-bosons obtain an effective mass is 
presented in~\cite{heavybaryons} in the case of 
description of baryons using heavy-flavor QCD. 

Examples of perturbative and TD approximations 
described above and in Secs. ~\ref{weakg}
and \ref{TD} in the exactly solvable model can be 
used in assessing convergence of similar approximations
in more complex cases. Consider the numerical studies 
of eigenvalue problems for TD Hamiltonian matrices 
obtained using bare quanta in the Yukawa and 
Yukawa-like theories, such as reported in~\cite{Li,Karmanov} 
and references therein. The same theories can be considered 
in the limit of fermion and boson masses much larger than 
the cutoff parameters, irrespective of the form of regularization. 
One can limit numerical calculations, where the quantum
degrees of freedom are discrete, to a single mode for 
all quanta involved, precisely as it is done here in the 
Yukawa theory to obtain our model Hamiltonian. In that 
setup, the TD Hamiltonian matrices one would obtain 
would resemble the ones in our model. One can compare 
the accuracy and convergence measures adopted 
in~\cite{Li,Karmanov} with exact results shown in 
Fig.~\ref{fig:TDillustration}. 

In our model case, the bare coupling constant $g = 
e g_0$, see Eq.~(\ref{gt}), determines the expectation 
value for the number of bare bosons, $\langle N_0\rangle$, 
in the exact fermion eigenstate. For $g$ small, a small 
$\langle N_0\rangle$ is expected. However, for $g$ 
order $\sqrt{4\pi}\sim3.5$, which corresponds to the 
conventional coupling constant $g^2/(4\pi) \sim 1$, 
Fig.~\ref{fig:TDillustration} shows that the expected 
number of bare bosons exceeds 30. It is stated
in~\cite{Li,Karmanov} that in theories considered there 
one achieves convergences using TD matrices with 3 
or 4 bosons for quite large coupling constants. It would 
hence be of interest to find out what mechanism is at 
work by which the inclusion of additional interactions 
and motion of bare bosons with respect to bare fermions 
improves the convergence so significantly. Convergence 
for the electromagnetic form factors may be less indicative 
of the number of bosons needed because the contributions 
of the Fock components with $n$ constituents at large 
momentum transfers may quickly decrease with $n$~\cite{BF}.

The fact that the TD matrix eigenvalue problems are
particularly suitable as a tool for seeking approximate
solutions to QFT in the FF of Hamiltonian dynamics  
originates in the special circumstance that the momentum 
component $p^+$ is conserved by the interactions and
cannot be negative, in a sharp distinction from the 
momentum component $p^z$ in the IF of dynamics,
which can have both signs. As a result, discretization
of momenta in a box on a front divides the available 
total momentum $P^+$ of an eigenstate of a FF
Hamiltonian into a definite number of pieces, say 
$K$. Each Fock-space constituent of an eigenstate 
must carry a natural number of units $P^+/K$. 
Therefore, the number of constituents is limited from 
above by $K$, which ties the maximal number of 
constituents in the TD approximation to the resolution
of momentum discretization, $K$. This is the basis of 
the so-called discretized light-cone quantization 
(DLCQ)~\cite{PB1,PB2,BPPreview}. 

The DLCQ method has been applied to the Yukawa 
theory~\cite{BrodskyHillerMcCartor1}. Divergences 
were regulated using the Pauli-Villars method that 
introduces additional massive fields. To obtain 
solvable models, the masses of quanta of the additional 
fields were set equal to those of the corresponding
physical ones~\cite{BrodskyHillerMcCartor2}. Similar 
DLCQ computations were also carried out in a solvable 
model that closely resembles the Yukawa theory 
of heavy fermions~\cite{GreenbergSchweber,Schweber}.
That model was used in~\cite{GreenbergSchweber} 
to introduce the concept of ``clothed'' particles. The 
clothed particle was defined using an exact solution for 
a state of a single particle. An analogous solution was 
recovered using DLCQ . In these examples, the DLCQ 
methods were found useful for constructing low-mass 
states in which the mean number of bare constituents 
was small.

As resulting from simplifications of one and the same
Yukawa theory, the applications of the DLCQ mentioned
above allow one to pin point basic features by which the 
RGPEP and DLCQ approaches differ. These features are 
visible in the three Eqs.~(\ref{Hmodel1}), (\ref{Hteff1}) 
and (\ref{Hinfty1}) of Sec.~\ref{spectrum}. They display 
three distinct operator forms of the same Hamiltonian 
that acts in the model Fock space. 

Equation~(\ref{Hmodel1}) corresponds to the initial, 
one might say, canonical Hamiltonian of a theory
without counter terms. This operator provides the 
starting point for the RGPEP, which is set up at the 
scale parameter $t = 0$. Since the model is ultraviolet 
finite, no ultraviolet divergences need to be countered.
 
Equation~(\ref{Hteff1}) displays the same Hamiltonian 
written in terms of creation and annihilation operators 
for the effective particles that correspond to an arbitrary 
positive value of the finite scale parameter $t$, as described 
in Sec.~\ref{EP}. The formula displays an effective fermion-boson 
interaction term and a fermion self-interaction term. The 
Hamiltonian has the universal form of a polynomial function 
of effective particle operators. The polynomial coefficients and 
operators are the computed functions of $t$. The self-interaction 
term vanishes at $t=0$ because there are no counter terms 
needed in the initial Hamiltonian. If instead the initial Hamiltonian 
led to divergences in any term of $H_t$ for any finite $t$, one 
would compute the counter terms at $t=0$ by demanding 
that the divergences in $H_t$ are eliminated. The model is 
too simple to illustrate in detail what is done in the RGPEP 
regarding computation of counter terms when the initial 
Hamiltonian is divergent. However, detailed perturbative 
illustrations are available in an asymptotically free example 
of a scalar theory in 5+1 dimensions~\cite{Glazek1} 
and in a general derivation of formulas for relativistic 
Hamiltonians of effective particles in QFT~\cite{Glazek2}. 
Here it is only noted that in the presence of divergent 
self-interactions, the self-interaction term would include a 
free, finite part of the corresponding counter term. That part 
would be adjusted by comparison with experiment and may 
be constrained by demands of symmetry that the resulting
theory is meant to posses.

Equation~(\ref{Hinfty1}) is an expression of the same model 
Hamiltonian in terms of the operators that create physical states
from the vacuum state. A state of a single physical particle is 
an eigenstate of the Hamiltonian. The formula~(\ref{Hinfty1}) 
is obtained in the limit $t \to \infty$. The effective creation 
and annihilation operators labeled by $\infty$ correspond to 
the physical particles of the model. Generally, $H_\infty$ that 
comes out of solving the RGPEP equation could involve mixing of 
eigenstates within degenerate multiplets that in addition to
the Hamiltonian eigenvalues are labeled by the eigenvalues of 
other operators that commute with $H_t$, such as a component 
of the angular momentum, spin, isospin or a similar quantity. 
Our model Hamiltonian form of Eq.~(\ref{Hinfty1}) corresponds 
to both the concept of ``clothed'' particles in~\cite{GreenbergSchweber} 
and the DLCQ solutions for single physical particle states. It is 
visible in Eq.~(\ref{Hinfty1}) that the model of Eq.~(\ref{Hmodel}) 
is too simple to produce interactions between the effective 
particles that correspond to $t=\infty$ and match physical 
ones as single-particle states. It is worth stressing that the 
effective particles for $t \to \infty$ do not have to correspond 
to the physical ones. This is important for considerations that 
involve the concept of confinement, see below and Sec.~\ref{c}.

It is now clear that the RGPEP and DLCQ computations 
discussed above differ significantly. The RGPEP produces 
a whole family of equivalent effective Hamiltonians. The 
DLCQ does not produce such a family. It does label 
Hamiltonian matrices with the resolution $K$ and the 
transverse momentum cutoff, introduced by the Pauli-Villars 
masses. However, these are the regularization parameters. 
The resolution $K$ and the Pauli-Villars masses are meant 
to be sent to infinity in order to obtain solutions of a theory. 
They are not the finite parameters analogous to the RGPEP 
$t$ on which the physical quantities do not depend, see 
Sec.~\ref{spectrum}. Each member of the family labeled 
by $t$ is expressed using a different choice of degrees of 
freedom in one and the same theory, which means using 
different creation and annihilation operators, or different 
quantum field operators that are built from them. Solving the 
TD Hamiltonian matrix eigenvalue problems in terms of bare 
quanta for which $t=0$ may be very difficult numerically 
because of involvement of many basis states in the dynamics, 
as is illustrated in the model by Fig.~\ref{fig:TDillustration}. 
An effective Hamiltonian with a finite $t$ that is adjusted to 
the scale of the physical quantity of interest is dominated by 
the effective basis states of a similar scale. An approximate 
but accurate description of the quantity of interest is simpler 
to achieve that way than by keeping all bare basis states in 
a computation that requires handling of all variables of the 
theory up to the cutoffs. The model example 
illustrates this feature solely in terms of the magnitude of 
the effective coupling constant that decreases as $t$ 
increases and thus weakens the coupling between different 
effective Fock components that correspond to the parameter
$t$. In contrast, the DLCQ approach attempts to solve the 
theory directly in terms of the degrees of freedom present 
in the quantum Hamiltonian in its initial form, analogous to 
Eq.~(\ref{Hmodel1}), {\it i.e.}, the one that is obtained by 
quantization of a local theory. 

Our model example makes it also clear that the concept 
of ``clothed'' particles mentioned above differs from the 
RGPEP concept of scale-dependent effective particles. 
The ``clothed'' particles approach is based on writing
a Hamiltonian in terms of operators associated with
the physical particles instead of the bare ones. In the 
RGPEP language, the idea is to replace the gradual 
evolution from $t=0$ to $t=\infty$ by a single jump. 
Such replacement is not available in any closed form 
in complex theories for which one does not have 
any exact solutions. Notably, in case of confinement
the required physical particles are not supposed to exist.
The issue is relevant to the ultimate DLCQ limit $K \to 
\infty$ that appears to be related to questions concerning 
the vacuum, which is assumed to carry $p^+=0$. The 
RGPEP approach is conceptually different from 
the ``clothed'' particle approach. Its equations can 
be solved for effective operators making various 
guesses or approximations and the resulting effective 
particles do not have to be identified with any physical, 
individually observable objects. The effective 
Hamiltonians $H_t$ can be studied in terms of their 
predictions for quantities accessible experimentally. 
For an example of such attempt in heavy-flavor 
QCD, see~\cite{heavybaryons}.

Another basic feature that distinguishes the RGPEP 
example from the DLCQ examples mentioned above 
is that the number of quantum degrees of freedom 
stays the same in the effective theory for all values 
of $t$, including the canonical theory at $t=0$. However, 
the interaction terms in $H_t$ evolve with $t$ as the 
RGPEP Eq.~(\ref{dHdt}) dictates. If the initial theory 
were divergent, the ultraviolet counter terms would 
be computed in the process of solving Eq.~(\ref{dHdt})
and they would be inserted in the initial condition at 
$t=0$. They would thus not be constructed by adding 
degrees of freedom like in the Pauli-Villars approach. 
Instead, the demand on the RGPEP evolution that for 
finite $t$ it yields finite effective Hamiltonians $H_t$ 
would be used to determine the missing counter terms 
in $H_0$.

Finally, it should be observed that in the non-relativistic
contexts of condensed matter physics, addressed broadly
in~\cite{Kehrein}, as well as in nuclear physics theory  
developed in~\cite{OSU} and elsewhere, similar Wegner-like 
equations and corresponding TD Hamiltonian matrix 
eigenvalue problems appear that resemble the ones 
obtained by applying the RGPEP to the model Hamiltonian 
of Eq.~(\ref{Hmodel}) or other model Hamiltonians of analogous 
nature, {\it cf.}~\cite{RomanianH}. According to the rule that 
the same equations have the same solutions, no matter 
what their interpretation is, and in view of the discussion of 
this section, it becomes clear that the RGPEP concept of 
effective particles developed in particle physics and explicitly 
illustrated using the elementary Eq.~(\ref{Hteff1}), can be 
introduced in the other branches of physical theory as well. 
For example, one can attempt to introduce a whole family 
of scale-dependent effective electron operators that include 
phonon operators in a model of a condensed-matter medium 
or effective nucleon operators that include meson operators 
in a model of a nucleus.

\section{Model solution and realistic theories}
\label{comparison}

The model solution illustrates the structure, function 
and purpose of the RGPEP in the context where no
divergences appear. The concept of counter terms 
only shows up through the cancellation of the fermion 
self-interaction energy, due to emission and absorption 
of bosons, against the effective fermion energy in the 
eigenvalue problem for the Hamiltonian $H_t$. The 
terms that cancel out are finite. The pattern is analogous 
to the cancellation between the finite parts of counter 
terms and self-interactions in realistic theories.

The model solution illustrates the weakening of effective
interactions solely in terms of the coupling constant that 
decreases as the RGPEP evolution parameter $t$ grows.
This weakening corresponds to the weakening obtained 
in terms of the vertex form factors in realistic theories.
The model running-coupling constant corresponds to the 
vertex form factor for the specific value of its argument, 
corresponding to the invariant mass change caused by
the interaction. Emergence of the RGPEP vertex form factors 
in the Yukawa theory is described in~\cite{GlazekWieckowski}. 
Analogous appearance of the vertex form factors in the 
Abelian gauge theory is shown in~\cite{Abelian}. The 
RGPEP form factors that emerge in the third-order 
computation of the effective vertices in a non-Abelian 
theory is provided in~\cite{GomezRocha}.

Extension of the model solution that would include the 
motion of bosons with respect to fermions and hence 
produce the associated vertex form factors, would be 
of great value. As pointed out earlier, the RGPEP vertex 
form factors are expected to be important in the derivation 
of effective quark and gluon dynamics in QCD. However, 
given the complexity of QCD, one might attempt to first 
undo some of the model simplifications made here and 
tackle the problem of applying the RGPEP to the Yukawa 
theory. To be specific, one may aim at a comprehensive 
resolution of the paradox that concerns interactions of 
nucleons with pions, and perhaps also other mesons. 
Namely, the exchange of just one pion between nucleons 
yields the Yukawa potential in second-order perturbtion 
theory, but the coupling constant one needs to introduce 
in order to match the phenomenology is so large that the 
standard perturbation theory with local interactions cannot 
be valid. Perhaps the large coupling corresponds not to a 
canonical Yukawa theory with $t=0$ but to the effective 
theory in which $t$ corresponds to the pion mass scale 
and the vertex form factors make the interaction effectively 
quite weak by suppressing it outside the small momentum 
transfer range that corresponds to the pion exchange. 

The model solution includes a coherent state of bosons 
around a fermion. One could ask if the pattern exhibited
by the model could be followed for the purpose of 
explaining if the effective Yukawa theory could describe 
the pion cloud around nucleons.

Since the RGPEP suppression of interactions corresponds 
to the vertex form factors, it makes sense to ask if any 
theory that introduces vertex form factors of some width 
might correspond to an effective one in the sense of the 
RGPEP for a width parameter $t$ matching the form factor 
scale. The example of particular interest is provided by 
the Nambu and Jona-Lasinio model~\cite{Nambu} that 
to the author's best knowledge was never analyzed using 
the RGPEP.

\section{Conclusion}
\label{c}

The main import of the elementary model study is that 
it illustrates how the RGPEP works in an exactly solvable model.
However, the realistic theories are much more  complex than the 
model and one cannot predict on the model basis if the 
RGPEP can fully provide the means that are required for 
unambiguous identification of the corresponding effective 
Hamiltonians in complex theories. To find out what can be 
achieved in that matter, one would have to focus on the 
direct application of the RGPEP in terms of perturbative 
expansions and TD approximations to the complex theories. 
In that context the model solution is of value because such 
approximate methods quickly get quite convoluted in realistic 
theories. The value is that one can use the model as a 
pattern to follow and to consult with when calculations 
get hard to see through. The key example of a barrier to 
break is to solve the RGPEP equation up to the fourth order 
of perturbation theory in QCD and derive the corresponding 
TD Hamiltonian matrices. Perhaps this is the way to obtain 
the constituent-quark picture of hadrons from QCD.

The case of quarks in QCD is particularly pressing even though 
one can also try to use the RGPEP for addressing theoretical 
issues of the Standard Model as a whole. The idea of constructing 
effective quarks dates back to early years of current 
algebra~\cite{Melosh}. As far as the author knows it is not fully 
realized till today, while the particle data tables~\cite{PDG} 
continue to classify hadrons mostly in terms of just two or 
three quark constituents. States that contain two more quarks 
are being added in the same spirit of constituents. 
QCD suggests instead that hadrons are built from practically 
unlimited numbers of quarks, antiquarks and gluons of canonical theory. 
Despite the great progress of lattice gauge theory, Gell-Mann's 
opinion from twenty years ago~\cite{Gell-Mann}  appears still 
valid: ``The mathematical consequences of QCD have still not 
been properly extracted, and so, although most of us are 
persuaded that it is the correct theory of hadronic phenomena, 
a really convincing proof still requires more work. It may be 
that it would be helpful to have some more satisfactory method 
of truncating the theory, say by means of collective coordinates, 
than is provided by the brute-force lattice gauge theory 
approximation!'' 

The author's opinion is that the basic difficulty to overcome 
before one can address precise phenomenology that involves  
fast moving and strongly interacting hadrons, is to first 
somehow gain control of the ground state of the theory. 
The reason is that all particle states one considers are 
meant to be created by action of operators on that special 
state. Such control is also desired concerning spontaneous 
breaking of symmetries. In the FF of Hamiltonian dynamics 
the vacuum problem is formulated in a different way than 
in the IF dynamics, {\it e.g.} see~\cite{KogutSusskind}. The condition 
$p^+ > 0$ for all quanta with finite momenta and non-zero 
masses can be compared with the condition that the vacuum 
state carries zero momentum. The vacuum state should also 
be invariant with respect to a change of an inertial frame of 
reference. This may be a large change, such as to the 
infinite moment frame used in the parton model. Somehow 
the vacuum state is limited to states akin to those with 
$p^+=0$, sometimes called the FF zero modes. 

The vacuum problem of QCD has a long history, stimulated 
by the concepts of quark and gluon condensates and 
posing questions in cosmology. To gain a perspective, one 
can consult the works~\cite{SVZ,Weinberg1,Weinberg2}. 
The leading condensates can be simply incorporated in the FF 
version of QCD sum rules~\cite{condensates} using the 
condition $p^+>\epsilon^+$ for all non-vacuum modes while 
the vacuum modes must have $p^+ < \epsilon^+$. The constant 
$\epsilon^+$ is treated as infinitesimal. If one assumed that 
the states with momenta $p^+ < \epsilon^+$were absent, one 
could even think that the cosmological vacuum problem may be 
resolved~\cite{BrodskyShrock}. However, the dynamics
of modes with $p^+ < \epsilon^+$ is singular and to
the author's best knowledge it is not understood.

Of course, the exact computation of effective Hamiltonians for 
the model of Eq.~(\ref{Hmodel}) is only relevant to the vacuum 
issue because the computation is used to illustrate the RGPEP. 
The point is that the vacuum problem in the FF Hamiltonians 
can be turned into a renormalization group issue according 
to~\cite{triangle}. Namely, the counter terms to the cutoff 
$\epsilon^+ \to 0$ are expected to mimic vacuum effects 
and one hopes to finesse dynamical effects due to the latter 
that way. The idea is presented in~\cite{triangle} using the 
FF power counting and original similarity renormalization 
group procedure~\cite{Similarity1,Similarity2}. However, 
the number and complexity of terms one obtains turns
out difficult to handle using the similarity procedure. With 
the RGPEP the situation is different because one does not 
need to directly address the multitude of matrix elements 
of many complex operators that involve initially unknown 
functions of many momentum variables. Therefore, one can 
focus instead on behavior of coefficients in polynomial functions 
of creation and annihilation operators for effective particles. 
Moreover, the RGPEP equation in QCD that corresponds to 
Eq.~(\ref{dHdt}) in our model discussion secures invariance 
of the Hamiltonians $H_t$ with respect to seven kinematical 
Poincar\'e symmetries, leaving only three that are dynamical 
and need to be renormalized. Consequently, instead of the 
cutoff $p^+ > \epsilon^+$ on the absolute momenta $p^+$, 
one can use a dimensionless cutoff $x > \epsilon$ on the 
ratio $x = p_1^+/p_2^+$ that momenta of particles 1 and 
2 involved in an interaction term can form. 

Exact non-perturbative solutions of the RGPEP equation in 
QFT as complex as QCD are not currently foreseeable. 
However, one can study the terms that emerge in 
perturbative expansion using asymptotic freedom,
known in the FF effective particle Hamiltonians to the 
lowest order only~\cite{GomezRocha}. The fourth-order 
calculation mentioned earlier is of interest because this is 
the first place where the running coupling appears in the 
effective interaction terms and increases with $t$. 
General fourth-order RGPEP formulas are available 
in~\cite{Glazek2}. As long as the effective coupling 
constant is not too large, one can use the perturbative 
expansion to learn what kinds of terms arise. Initial 
attempts at phenomenology included only second-order 
formulas for $H_t$ in QCD of heavy quarks~\cite{heavybaryons},
assuming that gluons gain effective masses. One needs to 
understand what happens in fourth order to see if there
are any signs of development of constituent quark masses 
for light quarks. If the masses emerge and the coupling
constant stays small enough, one can take advantage 
of the exact model and follow its pattern toward 
systematic improvement in accuracy of the computation.
 



\begin{thebibliography}{99}

\bibitem{Wilson1}
K. G. Wilson, 
Phys. Rev. {\bf 140}, B445 (1965).

\bibitem{triangle}
K. G. Wilson {\it et al.},
Phys. Rev. D {\bf 49}, 6720 (1994); Fig. 6.

\bibitem{HenleyThirring}
E. M. Henley, W. Thirring, 
{\it Elementary Quantum Field Theory}
(McGraw-Hill, New York, 1962).

\bibitem{Lee}
T. D. Lee,
Phys. Rev. {\bf 95}, 1329 (1954).


\bibitem{LMG}
H. J. Lipkin, N. Meshkov and A. J. Glick,
Nucl. Phys. {\bf 62}, 188 (1965).

\bibitem{Debergh}
N. Debergh and Fl. Stancu,
J. Phys. A: Math. Gen. {\bf 34}, 3265 (2001).

\bibitem{Providencia}
C. Provid\^encia, J. da Provid\^encia, Y. Tsue and M. Yamamura,
Progr. Theor. Phys. {\bf 116}, 87 (2006).

\bibitem{Louw}
J. C. Louw, J. N. Kriel and M. Kastner,
Phys. Rev. A {\bf 100}, 022115 (2019).

\bibitem{Dirac}
P. A. M. Dirac, 
Rev. Mod. Phys. {\bf 21}, 392 (1949).

\bibitem{GlazekPerry}
S. D. G{\l}azek and R. J. Perry,
Phys. Rev. D {\bf 45}, 3734 (1992).

\bibitem{algebra}
S. D. G{\l}azek and T. Mas{\l}owski,
Phys. Rev. D {\bf 65}, 065011 (2002).

\bibitem{Similarity1}
S. D. G{\l}azek, K. G. Wilson,
Phys. Rev. D {\bf 48}, 5863 (1993).

\bibitem{Similarity2}
S. D. G{\l}azek, K. G. Wilson,
Phys. Rev. D {\bf 49}, 4214 (1994).

\bibitem{Wegner}
F. Wegner, 
Ann. Phys. {\bf 506}, 77 (1994).

\bibitem{Kehrein}
S. Kehrein,
{\it The Flow Equation Approach to Many-Particle Systems},
Springer Tracts in Modern Physics, Vol. {\bf 217}, 2006.

\bibitem{Abelian}
S. D. G{\l}azek,
Phys. Rev. D {\bf 101}, 034005 (2020).

\bibitem{Tamm}
I. Tamm, J. Phys. (Moscow) {\bf 9}, 449 (1945).

\bibitem{Dancoff}
S. M. Dancoff, 
Phys. Rev. {\bf 78}, 382 (1950).

\bibitem{SzpigelPerry}
S. Szpigel and R. J. Perry, 
in {\it Quantum Field Theory, A 20th Century Profile}, 
ed. A. N. Mitra, p. 59; arXiv:hep-ph/0009071.



\bibitem{LFTammDancoff}
R. J. Perry, A. Harindranath, and K. G. Wilson, 
Phys. Rev. Lett. {\bf 65}, 2959 (1990).

\bibitem{PerryWilson}
R. J. Perry, K. G. Wilson,
Nucl. Phys. B {\bf 403}, 587 (1993).


\bibitem{Wilson2}
K. G. Wilson, 
Phys. Rev. D {\bf 2}, 1438 (1970).

\bibitem{heavybaryons}
K. Serafin {\it et al.},
Eur. Phys. J. C {\bf 78}, 964 (2018).

\bibitem{Li}
Y. Li, V. A. Karmanov, P. Maris, J. P. Vary, 
Phys. Lett. B {\bf 748}, 278 (2015).

\bibitem{Karmanov}
V. A. Karmanov, Y. Li, A. V. Simonov, J. P. Vary,
Phys. Rev. D {\bf 94}, 096008 (2016). 

\bibitem{BF}
S. J. Brodsky, G. R. Farrar,
Phys. Rev. D {\bf 11}, 1309 (1975).

\bibitem{PB1}
H. C. Pauli, S. J. Brodsky, 
Phys. Rev. D {\bf 32}, 1993 (1985).

\bibitem{PB2}
H. C. Pauli, S. J. Brodsky, 
Phys. Rev. D {\bf 32}, 2001 (1985).

\bibitem{BPPreview}
S. J. Brodsky, H-C. Pauli, S. S. Pinsky,
Phys. Rept. {\bf 301}, 299 (1998).

\bibitem{BrodskyHillerMcCartor1}
S. J. Brodsky, J. R. Hiller, G. McCartor,
Phys. Rev. D {\bf 58}, 025005 (1998).

\bibitem{BrodskyHillerMcCartor2}
S. J. Brodsky, J. R. Hiller, G. McCartor,
Ann. Phys. {\bf 296}, 406 (2002).

\bibitem{GreenbergSchweber}
O. W. Greenberg, S. S. Schweber, 
Nuovo Cimento 8, 378 (1958).

\bibitem{Schweber}
 S. S. Schweber, 
{\it An Introduction to Relativistic Quantum Field Theory}
(Row, Peterson and Co., Evanston, IL, 1961), p. 339.

\bibitem{Glazek1}
S. D. G{\l}azek,
Phys. Rev. D {\bf 60}, 105030 (1999). 

\bibitem{Glazek2}
S. D. G{\l}azek,
Acta Phys. Polon. B {\bf 43}, 1843 (2012).


\bibitem{OSU}
S. K. Bogner, R. J. Furnstahl, R. J. Perry,
Phys. Rev. C {\bf 75}, 061001 (2007).

\bibitem{RomanianH}
G. Ciobanu, V. B$\hat{\rm a}$rsan, A. T. Mincu,
Rom. Journ. Phys. {\bf 55}, 539 (2010).

\bibitem{GlazekWieckowski}
S. D. G{\l}azek and M. Wi\c eckowski,
Phys. Rev. D {\bf 66}, 016001 (2002).

\bibitem{GomezRocha}
M. Gómez-Rocha, S. D. Głazek,
Phys. Rev. D {\bf 92}, 065005 (2015).
 
\bibitem{Nambu}
Y. Nambu, G. Jona-Lasinio,
Phys. Rev. {\bf 124}, 246 (1961).

\bibitem{Melosh}
H. J. Melosh,
Phys. Rev. D {\bf 9}, 1095 (1973).

\bibitem{PDG}
P. A. Zyla {\it et al.} (Particle Data Group), 
Prog. Theor. Exp. Phys. 2020, 083C01 (2020).

\bibitem{Gell-Mann}
M. Gell-Mann, {\it Quarks, Color, and QCD}, in {\it The Rise of the Standard Model}, Eds. L. Hoddeson et al. (Cambridge University Press, 1999), 
p. 633.

\bibitem{KogutSusskind}
J. B. Kogut, L. Susskind,
Phys. Rept. {\bf 8}, 75 (1973).

\bibitem{SVZ}
M. A. Shifman, A. I. Vainshtein, V. A. Zakharov,
Nucl. Phys. B {\bf 147}, 385 (1979).

\bibitem{Weinberg1}
S. Weinberg,
Rev. Mod. Phys. {\bf 61}, 1 (1989).

\bibitem{Weinberg2}
S. Weinberg,
Phys. Rev. D {\bf 83}, 063508 (2011).

\bibitem{condensates}
S. D. G{\l}azek,
Phys. Rev. D {\bf 38}, 3277 (1988). 

\bibitem{BrodskyShrock}
S. J. Brodsky, R. Shrock,
Proc. Nat. Acad. Sci. {\bf 108}, 45 (2011).

\end{thebibliography}
\end{document}